\documentclass[11pt]{article}

\usepackage{epsfig}
\usepackage{amsmath}
\usepackage{amsthm}
\usepackage{amssymb}
\usepackage{amscd}

\chardef\bslash=`\\ % p. 424, TeXbook
\makeatletter
\def\verbatim{\interlinepenalty\@M \@verbatim
  \leftskip\@totalleftmargin\advance\leftskip2pc
  \frenchspacing\@vobeyspaces \@xverbatim} 
\makeatother
\numberwithin{equation}{section}

\def\1I{\relax{\rm 1\kern-.25em \rm l}} % Operator-Eins

\newcommand{\unity}{\1I}

\newcommand{\MyIm}{\Im\mathfrak{m}}
\newcommand{\MyRe}{\Re\mathfrak{e}}
\newcommand{\grad}{{\rm grad\,}}
\newcommand{\sgrad}{{\rm sgrad\,}}
\newcommand{\beqn}{\begin{eqnarray}}
\newcommand{\eeqn}{\end{eqnarray}}

\def\href#1#2{#2}

\font\tablefont=cmcsc10

\begin{document}

\thispagestyle{empty}
\rightline{HU-EP-01/11}
\rightline{hep-th/0103203}
\vspace{2truecm}
\centerline{\bf \Large Non-commutative D- and M-brane Bound States}
\vspace{1.5truecm}

\centerline{\bf \hspace{1ex}Boris K\"ors\footnote{koers@physik.hu-berlin.de},
Dieter L\"ust\footnote{luest@physik.hu-berlin.de}}
\vspace{-.2truecm}
\begin{center}
{\em 
      Institut f\"ur Physik \\
      Humboldt-Universit\"at \\ 
      D-10115 Berlin, Germany
}
\end{center}
\vspace{.3truecm}
\centerline{\bf and}
\vspace{.5truecm}
\centerline{\bf \hspace{1ex}Andr\'e Miemiec\footnote{a.miemiec@qmw.ac.uk}}
\vspace{-.2truecm}
\begin{center}
{\em
      Department of Physics \\ 
    Queen Mary and Westfield College \\
            Mile End Rd \\ 
       London E1 4NS, UK
}
\end{center}

\vspace{1.0truecm}
%%%%%%%%%%%%%%%%%%%%%%%%%%%%%%%%%%%%%%%%%%%%%%%%%%%%%%%%%%%%%
%      ABSTRACT
%%%%%%%%%%%%%%%%%%%%%%%%%%%%%%%%%%%%%%%%%%%%%%%%%%%%%%%%%%%%%
%\vspace{.4truecm}
\begin{abstract}
\noindent
We analyze certain brane bound states in M-theory and their descendants in
type IIA string theory, all involving 3-form or 2-form background fluxes.  
Among them are configurations which represent NCYM, NCOS and
OD$p$-theories in the scaling limit of OM-theory. In particular, we
show how the conditions for the embedding to preserve supersymmetry are
modified by the presence of the flux and discuss their relations for the
various different bound states. 
Via the formalism of geometric quantization
such a deformation of a supersymmetric cycle is related to a non-commutativity
of its coordinates.  We also study possible 
non-commutative deformations of the
Seiberg-Witten curve of ${\cal N}=2$ supersymmetric gauge theories
due to non-trivial $H$-flux.  
\end{abstract}
%%%%%%%%%%%%%%%%%%%%%%%%%%%%%%%%%%%%%%%%%%%%%%%%%%%%%%%%%%%%%
\bigskip \bigskip
\newpage

%\tableofcontents\newpage

%%%%%%%%%%%%%%%%%%%%%%%%%%%%%%%%%%%%%%%%%%%%%%%%%%%%%%%%%%%%%
%       SECTION:  Introduction
%%%%%%%%%%%%%%%%%%%%%%%%%%%%%%%%%%%%%%%%%%%%%%%%%%%%%%%%%%%%%
\section{Introduction}
\label{SectionIntro}

Non-commutative Yang-Mills theories (NCYM) arise as effective theories of
open strings whose endpoints move on D-branes in the background of a constant
antisymmetric tensor field $B_{ij}$ or magnetic field $F_{ij}$ 
on the branes \cite{Seiberg:1999vs}. The scaling limit which is involved to
decouple gravity and massive modes is schematically given by 
\begin{eqnarray*}
   \alpha' ~\rightarrow~ 0, \hspace{5ex} F_{ij}~\sim~ {\rm finite} .
\end{eqnarray*}
The S-duality of IIB superstrings maps the magnetic background
field $F_{ij}$ to electric fields $F_{0k}$,
\begin{eqnarray*}
F_{0k}~=~\frac{F_{ij}}{\sqrt{1\,+\,\left( F_{ij} \right)^2}}
  \, ,
\end{eqnarray*}
and 
the S-dual of NCYM is usually referred to as NCOS (non-commutative open
strings) \cite{Seiberg:2000ms,Gopakumar:2000na}. 
The above limit for $F_{ij}$ translates to a critical electric field 
which implies a theory of light open strings along the $0k$ directions and 
decoupled from gravity. 
In order to study non-perturbative configurations like supersymmetric 
instantons in NCYM, bound states of D0-D$(2p)$ branes with background magnetic
fields were considered, where supersymmetry puts some severe restrictions
on the form of the $F$-field. These systems are T-dual to D$p$-D$p'$ 
branes intersecting at certain angles which are likewise determined by
supersymmetry.\\ 

\noindent
In this paper we will discuss the lifting of such situations 
to M-theory and explain the relations to OM-theory (open membrane theory) 
\cite{Bergshoeff:2000jn,Gopakumar:2000ep}. 
Therefore we have to study non-threshold bound states of M5- and M2-branes,
which are equivalent to M5-branes in the presence of an antisymmetric 3-form
field $H_{ijk}$ living on the M5-brane world volume along the direction
of the M2-brane. 
The circle-compactification of the
M-theory set-up  involves the D4-D2 bound state, the
NS5-D2 and the D4-F1 bound states, 
and the scaling limit of OM-theory maps to the  
OD2-theory (open D2-brane theory) and to the NCOS limits, respectively. \\

\noindent
In the first part of this paper we will analyze the conditions on the
fluxes and the shape of these bound states in order to preserve
supersymmetry from a ten dimensional perspective. The conditions on 
the flux $F$ for the three types of bound states in type IIA
string theory
will be found equivalent to the self-duality condition of
$H$. In particular 
the supersymmetry
conditions on the D4-D2 brane bound state will be investigated
in some detail, where the D2-brane is rotated inside the D4-brane
by arbitrary angles and, in addition, magnetic fluxes on the D4-brane
are turned on. As a result we will show that the D2-brane in the
presence of the fluxes describes a deformed 2-cycle with an induced 
symplectic form determined by the flux. 
We then demonstrate by the techniques of geometric quantization 
that due to the non-trivial symplectic structure
the coordinates of the curve that describes the embedding of the D2-brane  
are non-commutative. \\

\noindent
 In the second part of the paper we will consider superpositions of
two  M5-branes, filling different spatial directions, 
or, in the smooth case, the 2-cycle which is formed by a single M5-brane.
The M5-branes will be bound
together with 
two different M2-brane configurations, namely first 
a separate M2-brane inside each M5-brane, corresponding
to M2-branes wrapping the 2-cycle, and second in a
fashion such that the intersection of any M2-brane with an M5-brane is a
self-dual string. The motivation for these configurations originates
from the observation \cite{Witten:1997sc}
that the embedding of a smooth M5-brane in M-theory  
for the case of vanishing background fields reproduces 
the Seiberg-Witten curve of ${\cal N}=2$ supersymmetric gauge theories.
Similarly the M-theory 
lift of certain ${\cal N}=1$  supersymmetric gauge theories 
leads to a supersymmetric M5-brane whose internal 
embedding is given by a supersymmetric 3-cycle, i.e. a special Lagrangian 
submanifold in 6-dimensions \cite{Karch:1999sj}.
We will see that for the first M2-brane configuration 
(see table~\ref{boundstate2})
the Seiberg-Witten curve will not be deformed by any of the allowed
fluxes, in agreement with some recent discussion on non-commutative
${\cal N}=2$ gauge theories \cite{Armoni:2001br}. \\

\noindent
 For the second M2-brane configuration (see table~\ref{braneconf2c}), 
following the work of
\cite{Gauntlett:1998vk,Gauntlett:1998wb,Gibbons:1998hm,Lambert:1998wc,Gutowski:1999tu},   
we have already considered the way a constant $H$-field affects the 
geometry of an M5-brane in \cite{Lust:1999pq} 
(see also \cite{Marino:2000af}). 
We will see that  
turning on $H$ indeed lifts the Lagrangian condition 
in the  BPS-equations of the M5-brane, which means that the Seiberg-Witten
curve in the presence of the flux gets deformed, such that 
it is no longer holomorphic in the same complex structure. 
Via geometric quantization this again signals a non-commutativity
on the deformed curve. But in the two dimensional case the holomorphicity 
can be restored by a rotation of coordinates in the definition of the 
complex structure.\\

%%%%%%%%%%%%%%%%%%%%%%%%%%%%%%%%%%%%%%%%%%%%%%%%%%%%%%%%%%%%%
%       SECTION:  
%%%%%%%%%%%%%%%%%%%%%%%%%%%%%%%%%%%%%%%%%%%%%%%%%%%%%%%%%%%%%

\section{Supersymmetric bound states} 
\label{secboundstates}

We first discuss a bound state of a single M5-brane with another single
M2-brane inside the M5-brane. From the M5 world volume perspective,
an M5-brane with a constant $H$-field represents a non-threshold
bound state of this M5-brane with an M2-brane along two spatial directions
of the M5-brane.
This will be the prototype and building block of
more complicated superpositions we shall encounter later on. \\

\subsection{M5-M2 bound states: OM-theory}
\label{SectionBPS}

In static gauge the embedding of the M5-brane into flat 11d spacetime 
is realized by a map, which describes the dependence of the 
transverse coordinates $X_i$, $i=6,\,{\ldots}\,,\,10$ on the
brane coordinates $x_j$, $j=0,\,{\ldots}\,,\,5$. Furthermore there 
is a two-form potential $B_{\mu \nu}$ 
on the six-dimensional worldvolume of the M5-brane with field 
strength $H=dB$. A nonlinear self duality constraint is realized
on the field $H$, so that the anti-self-dual part can be computed 
from the self-dual one:\\[-3ex] 
\begin{eqnarray*}
     H_{ijk}~=~\frac{4}{Q}\left(\,h_{ijk}\,+\,2\,(kh)_{ijk}\,\right) ,\\[-3ex]
\end{eqnarray*}
where $k_i^j$ and $Q$ are defined by $k_i^j=h_{ikl}h^{jkl}$ and
$Q=1-\frac{2}{3} {\rm tr} k^2$. If one considers the case with only 
$H_{012}$ and $H_{345}$ turned on, this constraint can be solved by 
parametrizing $H_{012}$ and $H_{345}$ through a single variable, which 
by convention is chosen to be 
\begin{eqnarray}\label{hselfdual}
   H_{012} ~=~ -\,\sin (\tilde\varphi),\hspace{6ex} 
   H_{345} ~=~    \tan (\tilde\varphi). 
\end{eqnarray}
Equivalently, this can be expressed as
\begin{eqnarray}\label{hselfduala}
   H_{012} ~=~ -\,\frac{H_{345}}{\sqrt{1\,+\,(H_{345})^2}}\, .
\end{eqnarray} 
The equations of motion of the M5-brane are obtained from the 
superembedding approach \cite{Howe:1997fb}. For the brane  
considered here the supersymmetry projector  $\,1-\Gamma\,$ 
is given by\\[-3ex] 
\begin{eqnarray} \label{BPS}
    \Gamma ~=~ \frac{1}{\sqrt{1+H_{ijk}^2}} \Gamma_{012345}
               ~-~ \frac{H_{ijk}}{\sqrt{1+H_{ijk}^2}}\, \Gamma_{ijk}.
\end{eqnarray}

\noindent
We now consider a bound state of an M5- and an M2-brane 
along the 012 directions of the world volume of
the M5-brane. This is the configuration of table~\ref{boundstate}. 
It will be of interest in the decoupling limits of
OM-theory, NCOS and OD2-theory, to be explained in the following section. \\  
%%%%%%%%%%%%%%%%%%%%%%%%%%%%%%%%%%%%%%%%%%%%%%%%%%%%%%%%%%%%%%%%%
%   TABLE: H-Field on 2 Cycle 
%%%%%%%%%%%%%%%%%%%%%%%%%%%%%%%%%%%%%%%%%%%%%%%%%%%%%%%%%%%%%%%%%
\parbox{\textwidth}
{
 \refstepcounter{table}
 \label{boundstate}
 \begin{center}
 \begin{tabular}{|c|ccccc|}
\hline
                      &   &   &   &   &      \\[-1.75ex]
  M5                  & 1 & 2 & 3 & 4 & 5    \\
  M2                  & 1 & 2 &   &   &      \\[1ex] 
 \hline
 \end{tabular} 
 \end{center}
 \center{{\tablefont Table~{\thetable}.} M5-M2 bound state configuration}
}\\[2ex]
%%%%%%%%%%%%%%%%%%%%%%%%%%%%%%%%%%%%%%%%%%%%%%%%%%%%%%%%%%%%%%%%

\noindent
This set-up has been discussed e.g. in \cite{Townsend:1997wg} and an according 
gravity solution was constructed in \cite{Izquierdo:1996ms}. One has to 
take into account
 that the projection operators $\Gamma_{012345}$ and $\Gamma_{012}$ 
no longer commute, such that the supersymmetric configuration must be a 
non-threshold BPS bound state. The conserved supersymmetry is determined by 
the projector
\begin{eqnarray} \label{proj}
  \Gamma^{(12)} ~=~ \cos(\tilde\varphi)\;\Gamma_{012345} 
                ~+~ \sin(\tilde\varphi)\;\Gamma_{012}
                ~=~ \Gamma_{012345}\,e^{\tilde\varphi\,\Gamma_{345}}.
\end{eqnarray}
It is equivalent to the projector of a single M5-brane with flux given by
$H_{012}= -\sin (\tilde\varphi)$ as in (\ref{BPS}). \\

\noindent
We would now like to address the issue of scaling limits which decouple
gravity from the theory on the M5-brane and  
conserves the influence of the flux $H_{012}$: OM-theory 
\cite{Gopakumar:2000ep} (see also \cite{Harmark:2001ff}). 
While the Planck mass $M_{\rm p}$ has to 
go to infinity, 
the effective scale $M_{\rm eff}$ of a membrane stretching along the 
worldvolume directions 012 of the M5-brane, which support the flux, can stay 
finite
\begin{eqnarray*}
    M^3_{\rm p} ~-~ \epsilon^{012}\,H_{012} ~=~ M_{\rm eff}^3 
                                           ~\sim~ {\rm finite} 
\end{eqnarray*}
by letting the flux go to its critical value, which implies: 
\begin{eqnarray*}
       H_{012} ~\sim~ M_{\rm p}^3 .
\end{eqnarray*}
The resulting theory has light excitations of M2-branes extending in the 012
directions of the M5-brane as dominating degrees of freedom. It has been
conjectured that the coordinates of the fluctuating 
M2-brane are non-commutative, but no proper derivation could be presented up
to now. \\

\noindent
Upon compactification on circles one can now recover supersymmetric bound 
states of type IIA branes. 
The M-theory-IIA dictionary then translates M5-branes to D4-branes or 
NS5-branes and M2-branes to D2-branes or fundamental F1 strings, 
respectively.
One has three options to pick some direction as 
the eleventh:  
\begin{itemize}
  \item parallel to only the M5-brane \hspace{3.5ex}
        $\;\;\Rightarrow\;\;$ D4-D2 bound state (NCYM-theory),
  \item parallel to both of the two branes 
        $\;\;\Rightarrow\;\;$ D4-F1 bound state (NCOS-theory) or
  \item orthogonal to both of them \hspace{6ex}
        $\;\;\Rightarrow\;\;$ NS5-D2 bound state (OD2-theory).
\end{itemize} 
The 3-form flux becomes an electric 2-form on the D4-branes 
or a RR 3-form flux on the NS5-branes:
\begin{eqnarray*}
  R_i\,H_{ijk} ~=~ F_{jk}\,.
\end{eqnarray*}
Let us go through the different options case by case.

\subsection{D4-D2 bound states: NCYM}

Compactifying along any of the 345 directions leads to a 
D4(01234)-D2(012) state. 
In the appropriate scaling limit $\alpha' \rightarrow 0$ with $F_{34} =
R_5H_{345}$ kept finite the resulting theory is NCYM
in (4+1) dimensions \cite{Gopakumar:2000ep}. Closed string states as well as 
massive open string excitations decouple, and one is left with pure YM theory 
on a non-commutative space.   
The flat superposition of the two kinds of 
D-branes is non-supersymmetric. 
%reflected by 2 directions with mixed 
%DN-boundary conditions. 
But, in the presence of constant magnetic 
background fluxes together with non-trivial intersection
angles of the two branes, supersymmetry can be restored.
In the following we will derive the combined
supersymmetry conditions  
on the fluxes and on the angles from the projector conditions on the spinors.

\subsubsection{Projector relations}

To be slightly more generic we
now consider  D$p$-D$q$-brane bound states, which one can easily
specialize to the D4-D2 case. The coordinates are chosen such that the
D$p$-brane extends into $0,1, \dots ,p$ 
directions, the D$q$-brane into $0,1, \dots ,q$
directions, and $q\le p$. The D$q$-brane can then be rotated inside the world
volume of the D$p$-brane by any angle $\varphi_{ij}$ in a plane
spanned by $x_i$ and $x_j$, with $i\le q$ and $q<j\le p$. 
The projector that defines the condition for preserving supersymmetry reads
\cite{Berkooz:1996km}  
\begin{eqnarray} \label{susyangles} 
    \Gamma_{01\ldots p}\, \epsilon      \,=\, \tilde{\epsilon}
    \;\;\;\;\;\&\;\;\;\;\; 
    \Gamma_{01\ldots q}\, (R \epsilon)  \,=\, (R \tilde{\epsilon})
    \;\;\;\;\Rightarrow\;\;\;\; 
    \Gamma_{(q+1)\ldots p}\,R^2 \epsilon \,=\, \pm \epsilon ,
\end{eqnarray}
where $R$ denotes the rotation 
\begin{eqnarray}
    R ~=~ \exp\left( \varphi_{ij} \Sigma_{ij} \right) 
\end{eqnarray} 
of the spinors, $\epsilon$ for left-moving, $\tilde{\epsilon}$ for 
right-moving supersymmetries and normalizations chosen such that 
$\Sigma_{ij}\,=\,\Gamma_{ij}/2$ takes eigenvalues $\pm i/2$. 

\noindent
Now we also turn on also magnetic fluxes 
$F_{ij}=\tan(\tilde \varphi_{ij})$ on the D$p$-brane.
Then the supersymmetry condition is identical to the 
last equation of
(\ref{susyangles}),
where we have replaced $\varphi_{ij}$ in $R$ by $\tilde\varphi_{ij}$, i.e.
\begin{eqnarray}
    \tilde R ~=~ \exp\left( \tilde\varphi_{ij} \Sigma_{ij} \right) 
\end{eqnarray} 
In fact, 
due to the change of
commutation relations, this is now derived from an asymmetric rotation
\begin{eqnarray} \label{susyfluxrot1} 
    \Gamma_{01\ldots q}\,\epsilon    \,=\,\tilde{\epsilon}
    \;\;\;\;\;\&\;\;\;\;\;
    \Gamma_{01\ldots p}\,(\tilde R \epsilon)\,=\,
    (\tilde R^{-1}\tilde{\epsilon})
    \;\;\;\;\Rightarrow \;\;\;\; 
    \Gamma_{(q+1)\ldots p}\,\tilde R^2\epsilon\,=\, \pm \epsilon ,
\end{eqnarray}
treating left- and right-moving spinors with opposite phases. This agrees with
the fact that the boundary conditions for an open string in the presence of
background 2-form flux can be derived from an asymmetric rotation from pure
Neumann or Dirichlet boundary conditions \cite{Blumenhagen:2000}.\\ 
%In both
%cases the effects of the rotation (\ref{susyangles}) or the flux
%(\ref{susyfluxrot1}) can be removed by a change of
%coordinates. \\ 

\noindent
These conditions for D-branes at angles and D-branes with flux 
can now be lifted to eleven dimensions, where we restrict ourselves to branes
of dimensions less that six. The first case is of course trivial,
two D-branes intersecting at any relative angle lift to two M-branes
with the same relative angle. For the second, we use the 
ten-dimensional chirality projector $(1-\Gamma_{11})$ 
in order to rewrite the asymetrically 
rotated equation in (\ref{susyfluxrot1}) as a condition
on the 11-dimensional spinor $\eta = \epsilon\oplus\tilde{\epsilon}$ 
\begin{eqnarray}
  \Gamma_{0 \ldots p}\, \tilde R \left( 1+\Gamma_{11} \right)\, \eta\, =\, 
  \tilde R^{-1} \left( 1-\Gamma_{11} \right)\, \eta .
\end{eqnarray}
 From this we derive
\begin{eqnarray*}
   \left(\, 
            1\,+\,\beta\,-\,(1\,-\,\beta)\,\Gamma_{11}\,
   \right)\, \eta 
   &=& \Gamma_{0 \ldots p}\, 
       \left(\, 
            \left(
                   \tilde{R}^2+\tilde{R}^{-2} 
            \right) \,+\, 
            \left( 
                   \tilde{R}^{2}-\tilde{R}^{-2} 
            \right)\,\Gamma_{11}\, 
       \right)\,\eta\nonumber \\ 
   &=& 2\,\Gamma_{0\ldots p}\, 
       \left( 
               \cos(\tilde\varphi_{ij}) 
              +\sin(\tilde\varphi_{ij})\,\Gamma_{ij}\Gamma_{11} 
       \right)\,\eta,  
\end{eqnarray*}
where we have used $\beta = \left(\Gamma_{0\ldots p}\right)^2 =\pm 1$. 
Obviously, asymmetric rotations or 2-form fluxes on D-branes are lifted to
become 3-form fluxes in M-theory. On the contrary, a similar computation 
shows that 
the conditions (\ref{susyangles}) lift trivially to 
eleven dimensions, angles indeed remain angles. All this is in accord with the
fact that in M-theory the 2-form field $F_{ij}$ is lifted to $H_{ij11}$, 
\begin{eqnarray*}
   R_{11}\,H_{ij11} ~=~ F_{ij}  .
\end{eqnarray*}
We have to expect that D-branes with 2-form flux lift to M-branes with
3-form flux simply because there is no 2-form present in M-theory. \\

\noindent
Upon compactification on a torus $\mathbb{T}^2$ the Dirac quantization requires
the flux to take discrete values only. Concretely, $F_{ij}$ is rational 
\begin{eqnarray} \label{fluxquant}
    F_{ij}~=~\frac{n}{m}\,\in\,\mathbb{Q}, \quad {\rm gcd}(m,n)=1 ,
\end{eqnarray}
where we have set the compact volume to one. The two integers $m$ and $n$ are
the D4- and D2-brane charges of the bound state. In the T-dual picture with
D2-branes at angles, they translate into the wrapping numbers of these D2-brane 
on the 1-cycles of the tori \cite{Blumenhagen:2000}. \\

\subsubsection{Bound States with supersymmetric field strength}

For simplicity consider first a D4-D2 bound state with vanishing angles
$\varphi_{ij}$ and only non-zero
fluxes with the  components $F_{12}=\tan(\tilde\varphi_{12})$ and 
$F_{34}=\tan(\tilde\varphi_{34})$.
If there is no 
flux in the 12 directions, $F_{12}=0$, one can employ T-duality to get 
the D2(034)-D0(0) bound state recently considered by \cite{Witten:2000mf}. 
An alternative description was presented in \cite{Blumenhagen:2000eb}, 
where another T-duality towards a D1(03)-D1(04') bound state of two 
D1-branes at a relative angle $\phi_{34}=\pi/2-\arctan(F_{34})$ 
in the 34-plane was used\footnote
%%%%%%%%%%%%%%%%%%%%%%%%%%%%%%%%%%%%%%%%%%%%%%%%%%%%%%%%%%%%%%%%%%%%%%%%%
%  FOOTNOTE
%%%%%%%%%%%%%%%%%%%%%%%%%%%%%%%%%%%%%%%%%%%%%%%%%%%%%%%%%%%%%%%%%%%%%%%%%
{
   In order not to confuse the notation, remember the $\varphi$'s denote  
   the rotation angles of the D2-brane inside the D4-brane; on the other 
   hand, the $\phi$'s are angles between the D-branes in the T-dual 
   picture. Finally the `angles' $\tilde\varphi$ always parametrize the 
   magnetic fluxes.
}.
%%%%%%%%%%%%%%%%%%%%%%%%%%%%%%%%%%%%%%%%%%%%%%%%%%%%%%%%%%%%%%%%%%%%%%%%
The condition for the system to preserve some 
supersymmetry is simply $\phi_{34}=0$, i.e.
$\pi/2=\arctan(F_{34})$, which translates into an 
infinite magnetic field $F_{34}$ on the D2(034)-brane. 
In fact, this configuration then is parallel and trivial. A more generic 
solution for the D4(01234)-D2(012) system to be supersymmetric is obtained 
by including relative flux $F_{12}$ as well. The 
configuration can then be T-dualized into a D2(013)-D2(02'4') state at 
relative angles 
$\phi_{12}=\tilde\varphi_{12}=\arctan(F_{12})$ and $\phi_{34}=
\pi/2-\tilde\varphi_{34}=\pi/2-\arctan(F_{34})$, 
primes indicating rotated 
directions.
Then
supersymmetry provides  the condition 
\begin{eqnarray} \label{fsusy}
     F_{12}~=~ -\frac{1}{F_{34}} \, ,
\end{eqnarray}
which translates via $F'_{12}\,=\,F_{12}$ and $F'_{34}\,=\,-1/F_{34}$ 
into the familiar self-duality condition $\,\ast F'\,=\,F'\,$ of the 
field strength $F'$ on a T-dual D0-D4 bound state. The non-commutative
deformation of the world volume of the D4-brane is ruled by the deformation
parameter 
\begin{eqnarray}
   \Theta_{12} \,=\, \frac{1}{F_{12}'} 
               \,=\, -\,\frac{1}{F_{34}'} 
               \,=\, -\Theta_{34} .
\end{eqnarray}
In M-theory the magnetic flux is of course lifted as
\begin{eqnarray}
R_5H_{345}=\tan(\tilde\varphi_{34})\, ,
\end{eqnarray}
in agreement with the parametrization in eq.~(\ref{hselfdual}).
An advantage of the D4-D2 bound state as compared to the M5-M2 state is that
we have got a completely well defined microscopic description in terms of open
string theory for this state. We can use our knowledge about such brane
configurations in type IIA to derive some properties of the M-brane state,
such as the self-duality (\ref{hselfdual}) and the projector conditions
(\ref{proj}) for preserving supersymmetry in the following.

\subsubsection{Supersymmetric cycles with flux} 

Actually, for the D4(01234)-D2(012) bound state from above 
there are more possibilities to be supersymmetric than just 
the flux $F_{12}=-1/F_{34}$. In principle we 
can introduce any kind of additional constant flux $F_{13,}, F_{24}, F_{14},
F_{23}$ on the D4-brane and 
rotate the D2-brane within the D4-brane by relative angles $\varphi_{13},\ 
\varphi_{14},\ \varphi_{23},\ \varphi_{24}$. These parameters describe 
a flat embedding, 
$f:\,(x_3,x_4)\rightarrow (\,x_3+iX_1(x_3,x_4),\,x_4+iX_2(x_3,x_4)\,)$, 
of the D2-brane  into the D4-brane, which in the presence of 
additional flux will not be a supersymmetric cycle anymore. 
In \cite{Blumenhagen:2000eb} the global conditions on angles 
to preserve any supersymmetry for flat D-branes were generalized to local
conditions on branes with non-trivial embedding into
space-time. This passing from global to local conditions implies a repacement 
of the so far flat brane by a generic supersymmetric cycle. The condition 
are usually phrased in terms of the pull backs of the
holomorphic 2-form $\Omega$ and of the symplectic form $\omega$ 
(see appendix \ref{AppendixNotation}),
defined for the 2-cycle in question by 
\begin{eqnarray} \label{notation}
f^\ast\omega     &=& \left(
                                  \partial_4 X_1
                                 -\partial_3 X_2
                         \right)\,
                         dx^3\wedge dx^4,\nonumber\\
f^\ast\MyRe\,\Omega &=& \left(
                                 1-
                                  \partial_3 X_1
                                  \partial_4 X_2  
                                 +\partial_4 X_1  
                                  \partial_3 X_2
                         \right)\,
                         dx^3 \wedge dx^4,\\
f^\ast\MyIm\,\Omega &=& \left(
                                  \partial_3 X_1                           
                                 +\partial_4 X_2
                         \right)\,
                         dx^3\wedge dx^4.\nonumber
\end{eqnarray}
%The induced metric on the brane, $g_{ij}$, fulfills a useful algebraic 
%identity 
%\begin{eqnarray}\label{identity}
%  \det\,g &=&  \left[f^\ast\MyIm\,\Omega\right]^2
%              +\left[f^\ast\MyRe\,\Omega\right]^2
%              +\left[f^\ast\omega\right]^2 . 
%\end{eqnarray}
%%%%%%%%%%%%%%%%%%%%%%%%%%%%%%%%%%%%%%%%%%%%%%%%%%%%%%%%%%%%%%%%%%%%%
The conditions for supersymmetry in the absence of fluxes
are then simply given by 
\begin{eqnarray} \label{fomnull}
    f^\ast\omega ~=~ f^\ast\MyIm\,\Omega ~=~ 0 .
\end{eqnarray}
In a similar vein we now combine the relations (\ref{susyfluxrot1}) and 
(\ref{susyangles}) for an asymmetric rotation $\tilde R_1$ and a symmetric 
rotation $R_2$ respectively to get the conditions for a cycle with flux\footnote
%%%%%%%%%%%%%%%%%%%%%%%%%%%%%%%%%%%%%%%%%%%%%%%%%%%%%%%%%%%%%
%  FOOTNOTE
%%%%%%%%%%%%%%%%%%%%%%%%%%%%%%%%%%%%%%%%%%%%%%%%%%%%%%%%%%%%% 
{
  These conditions can be analysed in full generality. 
}
%%%%%%%%%%%%%%%%%%%%%%%%%%%%%%%%%%%%%%%%%%%%%%%%%%%%%%%%%%%%%
\begin{eqnarray} \label{fluxandrot}
   \Gamma_{01234}\, (\tilde R_1 \epsilon)  \,=\,
   (\tilde R_1^{-1} \tilde{\epsilon})
   \;\;\;\;\&\;\;\;\;
   \Gamma_{012}\, (R_2 \epsilon)    \,=\,(R_2 \tilde{\epsilon})
   \;\;\;\Rightarrow\;\;\; 
   \Gamma_{34} \tilde R_1^2 R_2^2 ~\epsilon \,=\, \epsilon .
\end{eqnarray}
Now $\tilde{R}_1$ and $R_2$ are no longer commuting such that one cannot
remove the deformation by a change of coordinates, as was possible for
(\ref{susyangles}) and (\ref{susyfluxrot1}) separately.  
There are three sets of commuting rotations, those simultaneous 
in 12 and 34, those in 23 and
14 and those in 24 and 13 directions. By decomposing $\tilde{R}_1^2 R_2^2$ into
these three components we then get the conditions to preserve any supersymmetry: 
\begin{eqnarray} 
    0 &=& F_{12} \,+\, \frac{1}{F_{34}}, 
          \nonumber\\  
    0 &=& \varphi_{23}  \,-\, \varphi_{14} \,+\, 
          \arctan (F_{23})\,-\, \arctan(F_{14}) , 
          \label{susyfluxrot2}\\  
    0 &=& \varphi_{24}  \,+\, \varphi_{13} \,+\, 
          \arctan (F_{24})\,+\, \arctan(F_{13}).
          \nonumber 
\end{eqnarray}
Each line of (\ref{susyfluxrot2}) states a condition that is capable
to define a flat supersymmetric D-brane bound state by being satisfied
globally, whereas they may only be patched together locally. 
The three rotations, symmetric or asymmetric, corresponds to three different
relative $U(1)$
rotation of the two branes. Only together they generate the most general local
$SU(2)$ deformation of the globally flat cycle. One may simplify the
conditions by choosing coordinates where one of the $U(1)$ rotations is
absorbed, such that e.g. $f^\ast\MyIm\,\Omega=0$. 
The choice of relative signs in (\ref{susyfluxrot2}) 
is arbitrary and stems from a paricular 
choice of complex structure. It relates the equation with minus signs to 
the symplectic structure $f^\ast\omega$ and the one without to $f^\ast\MyIm\,
\Omega$. We now follow 
the usual procedure to replace the flat intersecting branes by a smooth curve
$X_1(x_3,x_4)$, $X_2(x_3,x_4)$, which means replacing
  the global angles $\varphi_{ij}$ by local quantities according to 
\begin{eqnarray}
     \tan (\varphi_{ij}) ~\hat{=}~ \partial_j X_i . 
\end{eqnarray}
Then we find 
\begin{eqnarray} \label{fomflux}
       \frac{f^\ast\omega}{f^\ast\MyRe\,\Omega}&=& -\,
       \frac{F_{23}\,-\,F_{14}}{1\,+\,F_{23}F_{14}}\, ,  \\
       \frac{f^\ast\MyIm\,\Omega}{f^\ast\MyRe\,\Omega} & =&- \, 
       \frac{F_{24}\,+\,F_{13}}{1\,-\,F_{24}F_{13}} \, .\nonumber  
\end{eqnarray}
These are the conditions that the deformed cycle preserves any 
supersymmetry in the presence of the 2-form flux on the D4-brane. 
Note, that in the case
$F_{23}=F_{14}=(*F)_{23}$ and $F_{24}=-F_{13}=(*F)_{24}$ the standard
conditions (\ref{fomnull}) are recovered. Then, the field strength and the
cycle are separately supersymmetric. In (\ref{fomflux}) the deviation of the
flux from being self-dual or anti-self-dual is compensated by the deviation of
the cycle from being special Lagrangian. \\

\noindent
The non-commutativity on the two-dimensional cycle, which describes the
embedding of the D2-brane into the D4-brane, due to the fluxes
$F_{23}$ and $F_{14}$ is now provided by the non-vanishing of
the induced symplectic form $f^\ast\omega$ in eq.~(\ref{fomflux}).
As explained in appendix \ref{AppendixGeomQuant} we can 
introduce a Poisson bracket via $f^\ast\omega$ and apply the formalism 
of geometric quantization to define commutator of the coordiates, i.e. 
\begin{eqnarray}
  \lbrack x^3, x^4\rbrack ~=~ i\, \{ x^3,x^4 \} 
                          ~=~ i\, 
                              f^\ast\omega (\,\sgrad x^3,\,\sgrad x^4\, ) .
\end{eqnarray}
Hence eq.~(\ref{fomflux}) suggests a
noncommutative deformation of the operators which are associated to the
coordinates of the cycle that describes the embedding of the
D2-brane into the D4-brane by the presence of additional 2-form flux. The flux
$F_{12}=-1/F_{34}$ would not have been sufficient to make this construction. 
In what sense this non-commutativity arising from (\ref{fomflux}) can
be understood from a microscopic point of view, and how the values of the
non-commutativity parameters can be reconciled with the $\Theta^{ij}$
parameters known for NCYM, remains an open question. \\

\noindent
Whenever the flux of the D4-D2 system is not tuned in the
supersymmetric fashion, there will appear a tachyon in the open string
spectrum of strings stretching between the two branes. It is believed that
this signals a condensation mechanism towards the true ground state of the
system, which is again BPS. 
At the critical point of the field strength, when the tachyon
becomes massless, one suspects a marginal deformation that takes the bound
state with tachyon condensate into the state described by (\ref{fomflux}). 
This is only a special case of the more general scenarios of
Dq-Dp bound states considered in 
\cite{Witten:2000mf,Blumenhagen:2000eb,Mihailescu:2000dn,Ohta:2001dh}. 
%Also,
%the D4-D2 system is only a special case of more general 
%D$p$-D($p/2$) bound states, where the fluxes lead to a deformation
%of a supersymmetric $p$-cycle, i.e. to a  deformation of a $p$-dimensional
%special Lagrangian submanifold in ${\mathbb R}^{2p}$.

\subsection{D4-F1 bound states: NCOS}

The M2-M5 configuration also allows compactification in 12 directions, say we
take 2, to a D4-F1 bound state. 
%This will lead to the (4+1)
%dimensional NCOS-theory.
It is related by a chain of a T-duality along any of the 345 directions, e.g. 5, the S-duality of type IIB and another T-duality 
along 5 to the D4-D2 bound state
discussed above \cite{Townsend:1997wg} after only exchanging the labels for 
the 2 and 5 directions (see also \cite{Russo:2000mg,Russo:2000zb}). 
The dualities also relate the conditions for
preserving supersymmetry by just erasing $\Gamma$'s and switching chiralities
appropriately. The mapping of the magnetic flux $F_{34}$ to an electric flux
$F_{01}$ has been given in \cite{Gopakumar:2000na} by 
\begin{eqnarray} \label{f01}
   F_{01} ~=~ \,\frac{F_{34}}{\sqrt{1\,+\,\left( F_{34} \right)^2}}
          ~=~ -\,\frac{1}{\sqrt{1\,+\,\left( F_{12} \right)^2}}
          ~=~ -\, \sin (\tilde\varphi) \, ,
\end{eqnarray}
via the condition eq.~(\ref{fsusy}) imposed by supersymmetry of the D4-D2
state  
on the $F$-flux. A formula for the Dirac quantization of 
the flux on a D$p$-F1 bound states has also been derived in
\cite{Lerda:1999} by boundary state techniques. It reproduces the flux quantization
(\ref{fluxquant}) on the D4-D2 bound state. Dual gravity solutions have been
constructed in \cite{Lu:1999,LaSu}. 
Lifting the flux to eleven dimension gives the electric component 
\begin{eqnarray*}
   R_2\,H_{012}~=~F_{01}~=~ -\,\sin (\tilde\varphi)\,
\end{eqnarray*}
of the 3-form. Together with (\ref{f01}) this precisely reproduces the 
condition (\ref{hselfdual}) 
on $H$ after scaling the radii to 1. In other words, the supersymmetry 
of the type IIA D4-D2 bound state implies the self-duality of $H$. \\ 

\noindent
It has been shown in \cite{Gopakumar:2000na} that the OM scaling limit of the
M5-M2 bound state reduces to the NCOS limit on a D4-brane provided the radius
of the circle and the flux is tuned in a particular way:
\begin{eqnarray} \label{scales}
   R_2 ~=~ G^2_{\rm o}\;\sqrt{\alpha'_{\rm eff}},\hspace{6ex} 
   M^3_{\rm eff} ~=~ \frac{1}{2\,G^2_{\rm o}\,\alpha^{\prime 3/2}_{\rm eff}}
   \, .
\end{eqnarray} 
Then $G_{\rm o}$ stands for the open string coupling and $\alpha'_{\rm eff}$
for the effective scale of fundamental strings. 
In NCOS the closed string excitations decouple from the brane but all open string states remain at finite mass. 
The electric field introduces a non-commutativity of the time-space
coodinates 01, governed by the parameter  
\beqn
\Theta_{01} = \Theta_{34} G^2_{\rm o}
\eeqn
This limit has been introduced as the weakly coupled S-dual of the strongly
coupled NCYM-theory that arises on a D-brane in the presence of a magnetic
2-form flux, when going to strong coupling. \\

\subsection{NS5-D2 bound states: OD2-theory}

By compactifying along any of transverse directions 678910 
one obtains a D2-brane inside an NS5-brane with a RR-flux $C_{ijk}$ inherited
from $H_{ijk}$. This state is related via a T-duality in 3 direction, an 
S-duality which trades the NS5-brane for a D5-brane and another T-duality on 3 
to the D4-D2 bound state discussed above, after only exchanging the 
labels 3 and 5. This chain of dualities transforms the 3-form $C_{345}$ on the 
NS5-brane into the magnetic flux $F_{34}$ on the D4-brane. 
In \cite{Mitra:2000wr,LaSu} the gravity solution for such a bound
state has been calculated. From the formulas for the 
RR-flux on the NS5-D2 bound state given there, one can read off 
the coefficients
\begin{eqnarray*}
   C_{012} ~=~ -\,\sin (\tilde\varphi) ,\hspace{6ex} C_{345} ~=~ 
   \tan(\tilde\varphi)
\end{eqnarray*}
The Dirac quantization again forces $\tan(\tilde\varphi)$ to be
rational. Together, the NS5-D2 bound state also carries the information of the
self-duality of $H$. \\

\noindent
By tuning the radius and the flux of the resultant OM-theory 
formally in the same way as in (\ref{scales}) 
\begin{eqnarray} \label{scales2}
  R_6 ~=~ G^2_{{\rm o}2}\,\sqrt{\tilde{\alpha}'_{\rm eff}},\hspace{6ex} 
  M^3_{\rm eff} ~=~ \frac{1}{2\,G^2_{{\rm o}2}\,\tilde{\alpha}^{\prime
  3/2}_{\rm eff}}\, ,
\end{eqnarray} 
one gets the so-called OD2-theory with 
light open D2-branes excitations on the world 
volume of the NS5-brane but closed strings decoupled again. 
$G_{{\rm o}2}$ and $\tilde{\alpha}'_{\rm eff}$ are now 
the D2-brane  coupling constant and the effective scale of fundamental
strings. \\

%%%%%%%%%%%%%%%%%%%%%%%%%%%%%%%%%%%%%%%%%%%%%%%%%%%%%%%%%%%%%%%%%%%%
% Section: Superpositions
%%%%%%%%%%%%%%%%%%%%%%%%%%%%%%%%%%%%%%%%%%%%%%%%%%%%%%%%%%%%%%%%%%%%

\section{Superpositions of M-brane bound states} 
\label{secseibergwitten}

In this section we study two types of superpositions of M2-M5 bound states. The
basic building block in both cases is the K\"ahler calibration of two
M5-branes as given in table~\ref{KaehlerCalibration}. \\
%%%%%%%%%%%%%%%%%%%%%%%%%%%%%%%%%%%%%%%%%%%%%%%%%%%%%%%%%%%%%%%%%
%   TABLE: H-Field on 2 Cycle 
%%%%%%%%%%%%%%%%%%%%%%%%%%%%%%%%%%%%%%%%%%%%%%%%%%%%%%%%%%%%%%%%%
\parbox{\textwidth}
{
 \refstepcounter{table}
 \label{KaehlerCalibration}
 \begin{center}
 \begin{tabular}{|c|ccccccc|}
 \hline
%                      &   &   &   &   &   &   &   \\[-1.75ex]
%                      & 1 & 2 &   &   &   & 6 & 7 \\[0.5ex]
% \hline
% \hline
                      &   &   &   &   &   &   &   \\[-1.5ex]
  M5                  & 1 & 2 & 3 & 4 & 5 &   &   \\
  M5'                 &   &   & 3 & 4 & 5 & 6 & 7 \\[0.5ex]
 \hline
 \end{tabular} 
 \end{center}
 \center{{\tablefont Table~{\thetable}.} K\"ahler calibration}
}\\[2ex]
%%%%%%%%%%%%%%%%%%%%%%%%%%%%%%%%%%%%%%%%%%%%%%%%%%%%%%%%%%%%%%%%
It consists of two intersecting M5-branes, whose
embedding into the $\mathbb{R}^4$ spanned by the 1267 directions is the
holomorphic cycle that governs the dynamics of the ${\cal N}=2$ gauge theory
in the four non-compact space-time directions 0345 \cite{Witten:1997sc}. 
In the following, this
setting is then combined with additional M2-branes. First, in the way of the
previous chapters, a separate M2-brane inside each M5-brane, and second in a
fashion such that the intersection of any M2-brane with an M5-brane is a
self-dual string. The first is a direct generalization of the previous
chapters. It will not lead to a deformation of the Seiberg-Witten curve,
whereas the second case does. 

\subsection{M2-branes wrapping the Seiberg-Witten curve}

Because the respective projectors (\ref{proj}) mutually commute 
the bound states that have been studied so far  
as isolated states can also be superposed:
\begin{eqnarray*}
   [\, \Gamma^{(12)},\, \Gamma^{(67)}\, ] ~=~0 .
\end{eqnarray*} 
Such a configuration is the set-up of table~\ref{boundstate2}. \\ 
%%%%%%%%%%%%%%%%%%%%%%%%%%%%%%%%%%%%%%%%%%%%%%%%%%%%%%%%%%%%%%%%%
%   TABLE: H-Field on 2 Cycle 
%%%%%%%%%%%%%%%%%%%%%%%%%%%%%%%%%%%%%%%%%%%%%%%%%%%%%%%%%%%%%%%%%
\parbox{\textwidth}
{
 \refstepcounter{table}
 \label{boundstate2}
 \begin{center}
 \begin{tabular}{|c|ccccccc|}
 \hline
%                      &   &   &   &   &   &   &   \\[-1.75ex]
 %                     & 1 & 2 &   &   &   & 6 & 7 \\[0.5ex]
% \hline
% \hline
                      &   &   &   &   &   &   &   \\[-1.5ex]
  M5                  & 1 & 2 & 3 & 4 & 5 &   &   \\
  M2                  & 1 & 2 &   &   &   &   &   \\
%                     &   &   &   &   &   &   &   \\
  M5'                 &   &   & 3 & 4 & 5 & 6 & 7 \\
  M2'                 &   &   &   &   &   & 6 & 7 \\[1ex]
 \hline
 \end{tabular} 
 \end{center}
 \center{{\tablefont Table~{\thetable}.} Bound state superposition}
}\\[2ex]
%%%%%%%%%%%%%%%%%%%%%%%%%%%%%%%%%%%%%%%%%%%%%%%%%%%%%%%%%%%%%%%%
It may be interpreted as an M2-brane
wrapping the Seiberg-Witten curve, or, equivalently, as a Seiberg-Witten curve
with additional 3-form flux turned on. \\

\noindent
In general, a compactification down to type IIA now leads to a
superposition of either NS5-D2 and D4-F1 or of two D4-D2 bound states. 
Along the compact eleventh direction the configuration
looks like F1 and D4, with 2-form flux on the D4 along the world volume of the
F1, and transverse to it, like D2 and NS5 with 3-form flux on the NS5 along the
worldvolume of the D2-brane, which has been depicted in 
figure~\ref{swcurve}. \\[2ex]
%%%%%%%%%%%%%%%%%%%%%%%%%%%%%%%%%%%%%%%%%%%%%%%%%%%%%
%   BILD: 
%%%%%%%%%%%%%%%%%%%%%%%%%%%%%%%%%%%%%%%%%%%%%%%%%%%%%
\parbox{\textwidth}
{ 
 \refstepcounter{figure}
 \label{swcurve}
 \begin{center}
 \makebox[10cm]{
  \epsfxsize=14cm
  \epsfysize=6cm
  \epsfbox{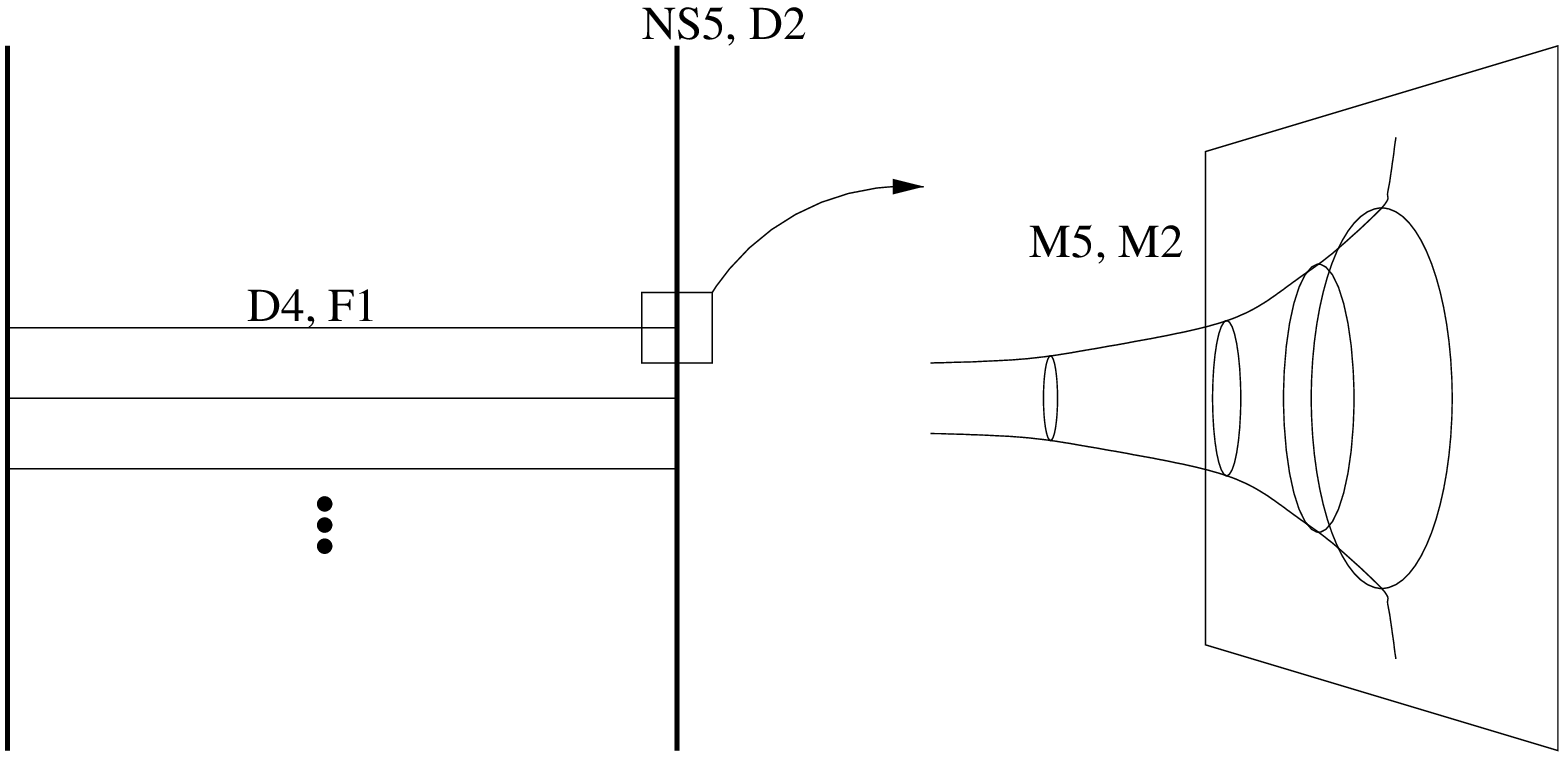}
 }
 \end{center}
 \center{{\bf Fig.~\thefigure} Blowing up the eleventh direction}
}\\[1ex]
%%%%%%%%%%%%%%%%%%%%%%%%%%%%%%%%%%%%%%%%%%%%%%%%%%%%%
%
%%%%%%%%%%%%%%%%%%%%%%%%%%%%%%%%%%%%%%%%%%%%%%%%%%%%%
The scaling limit of OM-theory now applied to the superpositions of 
bound states then reveals a superposition of such effective
theories of noncommutative open D2-branes and fundamental strings. 
Upon compactification along any of the 1267 directions we get NCOS along the
eleventh direction and OD2 transverse to it.
Interestingly, both, the NCOS and the OD2 scaling limits (\ref{scales}) and
(\ref{scales2}) are compatible:  
Given the radius $R_{11}$ and the critical scaling limit of the
flux, i.e. $M_{\rm eff}$, the couplings 
and, as well, the fundamental string scales are identical for both, the
OD2 fluctuations on the NS5-brane, as well as the open string fluctuations on
the D4-brane. This had to be expected as both stem from the fluctuations of
M2-branes in the elevendimensional OM-theory. 
While a complete description of the (NS5,D4,D2,F1) bound state and its scaling
limit is not known to us, 
partial results for such bound states have for instance been
discussed in \cite{Mitra:2000wr,Harmark:2000wv}. There gravity duals of 
bound states of the kind (NS5,D4,D2) and (D4,F1) have been
derived and the NCOS or OD2 limits of such systems have been analyzed. \\ 

\noindent
Let us again look closer at the limit which we know most about, a
compactification along the 5 direction towards a D4-D2-D4'-D2' bound
state. We call the field strength on the first D4-D2 bound state $F$, on the
second $F'$ and have the equations (\ref{fomflux}) for both states separately
after adapting indices. The Seiberg-Witten curve is obtained by analyzing the
embedding of $X_6$ and $X_7$ as functions of $x_1$ and $x_2$. If we now write
symmetric and asymmetric rotation operators for any kind of angle and flux
which can occur for the state of table~\ref{boundstate2}, we have to realize
that there are no fluxes that could deform the Seiberg-Witten curve. 
Deforming the
equations 
\beqn
\varphi_{16} + \varphi_{27} &=& 0\ \Rightarrow\ f^\ast\omega =0, \\ \nonumber
\varphi_{17} + \varphi_{26} &=& 0\ \Rightarrow\ f^\ast\MyIm\,\Omega =0 
\eeqn 
would require fluxes $F_{ij},\ i\in \{ 1,2\},\ j\in\{ 6,7\}$, which are
not present here. This situation is very similar to that of D$p$-D$p'$ bound
states in \cite{Blumenhagen:2000eb}. 
All the other equations which can be derived analogously to
(\ref{fomflux}) decouple, and are identical to those of an isolated
D4-D2 bound state. Thus, we have to conclude that the supersymmetric cycle,
which describes the embedding of the D4'-brane relative to the D4-brane is
not deformed and remains special Lagrangian. Such we have to expect 
that also in M-theory an M2-brane that wraps the Seiberg-Witten curve does not lead
to any non-commutative deformation of the curve, which should otherwise be
inherited by the type IIA realization. 
This result is supported by \cite{Armoni:2001br},
where non-commutative ${\cal N}=2$ gauge theories
were recently studied by more direct means 
with the result that the Seiberg-Witten
curve is kept unchanged.

\subsection{Self-dual strings on an M5-brane}
\label{SubSectionStringsOnM5}

Finally we like to consider another configuration of M5- and M2-branes, where
the M2-branes intersect the world volume of the M5-branes in self-dual
strings. It has been shown to describe ${\cal N}=2$ supersymmetric gauge
theory in the presence of BPS monopoles whose positions are given by the
positions of the M2-branes in the space-time directions 0345. \\

%%%%%%%%%%%%%%%%%%%%%%%%%%%%%%%%%%%%%%%%%%%%%%%%%%%%%%%%%%%%%%%%%
%   TABLE: H-Field on 2 Cycle 
%%%%%%%%%%%%%%%%%%%%%%%%%%%%%%%%%%%%%%%%%%%%%%%%%%%%%%%%%%%%%%%%%
\parbox{\textwidth}
{
 \refstepcounter{table}
 \label{braneconf2c}
 \begin{center}
 \begin{tabular}{|c|ccccccc|}
 \hline
%                      &   &   &   &   &   &   &   \\[-1.75ex]
%                      & 1 & 2 &   &   &   & 6& 7\\[0.5ex]
% \hline
% \hline
                      &   &   &   &   &   &   &   \\[-1.5ex]
  $\overline{\rm M5}$ & 1 & 2 & 3 & 4 & 5 &   &   \\
  M5                  &   &   & 3 & 4 & 5 & 6 & 7 \\
  M2                  & 1 &   &   &   &   & 6 &   \\
  M2                  &   & 2 &   &   &   &   & 7 \\[1ex]
% \hline
%  \vbox{\vspace{0.3ex}\hbox{$\overline{\rm M2}$}} & 0 &   & 2 &   &   &   & 6 &   \\
 \hline
 \end{tabular} 
 \end{center}
 \center{{\tablefont Table~{\thetable}.} Brane configuration}
}\\[2ex]
%%%%%%%%%%%%%%%%%%%%%%%%%%%%%%%%%%%%%%%%%%%%%%%%%%%%%%%%%%%%%%%%
\noindent
Now the M2-brane end as points on either of the two asymptotic M5-branes.
We want to study again the embedding of the two-dimensional curve
in the directions 1267, which is described by the embedding map  
$f:\,(x_1,x_2)\rightarrow (\,x_1+iX_6(x_1,x_2),\,x_2+iX_7(x_1,x_2)\,)$. 
Unlike the previous M5-M5' bound state the 2-cycle will be now deformed
by the presence of the $H$-field. The derivation of the BPS-equations 
is identical to the derivation in \cite{Lust:1999pq} or the more general 
discussion in \cite{GauntlettMemOnFive}. 
The tangent frame the non zero components of $H$ can also be expressed 
in terms of $h_{012}$. The two relevant components of eq.~(\ref{hselfdual}) 
read:\\
\begin{eqnarray}\label{EqKomponeten}
   H_{012} ~=~ -\,\sin\tilde\varphi 
           ~=~ \frac{4\,h_{012}}{4\,h_{012}^2+1} , \hspace{5ex}
   H_{345} ~=~    \tan\tilde\varphi 
           ~=~ \frac{4\,h_{012}}{4\,h_{012}^2-1}.
\end{eqnarray}
The spacetime frame components are distinguished from them by an 
additional twiddle $\tilde{H}_{012}$.
The BPS-equations (compare with eq.~\ref{fomflux}) read
\begin{eqnarray}\label{Symplectic}
    \left[f^\ast\omega\right] 
    &=& \,\tilde{H}_{012}
    ~=~ -\;\sin\tilde\varphi\,\sqrt{-g} , \\[1ex] 
    \left[f^\ast\MyRe\,\Omega\right] 
    &=& \,\frac{\tilde{H}_{012}}{\tilde{H}_{345}}
    ~=~ -\,\cos\tilde\varphi\,\sqrt{-g} , \label{Volume}\\[1ex]
    \left[f^\ast\MyIm\,\Omega\right] &=& 0 . \label{ImagTeil}
\end{eqnarray}
Equation~(\ref{Symplectic}) is a gauged symplectic structure, i.e. it is 
proportional to $H_{012}$. 
Again, using the methods of geometric quantization, it leads to a
non-commutativity on the deformed cycle.
The tangent frame component $H_{012}$
can be easily interpreted with respect to the Grassmanian
$G(2,4) = S^2_{+}\times S^2_{-}$, defined in appendix~\ref{AppendixNotation}. 
The algebraic identity  
\begin{eqnarray}\label{identity}
  \det\,g &=&  \left[f^\ast\MyIm\,\Omega\right]^2
              +\left[f^\ast\MyRe\,\Omega\right]^2
             +\left[f^\ast\omega\right]^2 . 
\end{eqnarray}
provides a parametrisation of the sphere $S^2_{+}$ by: 
\begin{eqnarray}
  -\,[f^\ast\omega]/\sqrt{g}        &=& \sin\tilde\vartheta\,
                                        \sin\tilde\varphi , \nonumber\\{}
  -\,[f^\ast\MyRe\,\Omega]/\sqrt{g} &=& \sin\tilde\vartheta\,
                                        \cos\tilde\varphi , 
                                        \label{Parametrisation}\\{}
  -\,[f^\ast\MyIm\,\Omega]/\sqrt{g} &=& \cos\tilde\vartheta . \nonumber 
\end{eqnarray}
Obviously eq.~(\ref{ImagTeil}) leads to $\tilde\vartheta=\pi/2$ and 
we can identify the component $H_{012}$ with an angle of 
$S^2_{+}$ by:  
\begin{eqnarray*}
      H_{012} &=& -\,\sin\tilde\varphi . 
\end{eqnarray*}
As explained at the end of appendix~\ref{AppendixNotation} this 
can be interpreted as a brane rotation if one compares with the 
standard choice of holomorphic coordinates.\\

%%%%%%%%%%%%%%%%%%%%%%%%%%%%%%%%%%%%%%%%%%%%%%%%%%%%%%%%%%%%%
% Section: Appendix
%%%%%%%%%%%%%%%%%%%%%%%%%%%%%%%%%%%%%%%%%%%%%%%%%%%%%%%%%%%%%

\begin{appendix}

\section{Notation}
\label{AppendixNotation}

The definition of the BPS-solutions can be restated in terms of  
certain closed p-forms (calibrations), which are defined after fixing 
a complex structure on the embedding space. Since we are mainly 
concerned with 2-cycles in ${\mathbb{R}}^4$, the space of 
$q^1,p^1,q^2,p^2$, the set of complex 
structures is given by the set of matrices 
\begin{eqnarray*}
      J &=& \left(
                \begin{array}{cccc}
                    ~0    & -a_{1} & -a_{2} & -a_{3} \\
                   a_{1} &   ~0    & -a_{3} & \hspace{1.7ex}a_{2} \\
                   a_{2} & \hspace{1.7ex}a_{3} &    ~0   & -a_{1} \\
                   a_{3} & -a_{2} & \hspace{1.7ex}a_{1} &  ~0      
                \end{array}
            \right) 
\end{eqnarray*}
with $(a_1,a_2,a_3)$ a point on a sphere of unit length. 
If we select one of them, say \hbox{$J_0\,=\,(1,0,0)$}, 
the complex coordinates are \hbox{$z^1=q^1\,+\,i\,p^1$} and 
\hbox{$z^2=q^2\,+\,i\,p^2$}.\\
In these complex coordinates  the sphere of complex structures may be  
identified with the sphere of selfdual 2-forms in ${\mathbb{R}}^4$. 
Then the coordinates $(a_1,a_2,a_3)$ refer to the basis below:  
\begin{eqnarray}
        \omega &=& \frac{i}{2}\left(\,
                                       dz^1\wedge d\bar{z}^1 ~+~
                                       dz^2\wedge d\bar{z}^2
                              \right) , \nonumber\\
        \MyRe\,\Omega &=& \;\MyRe\,\left(\,dz^1\wedge dz^2 \right) , 
         \label{BasisSelfdual}\\
        \MyIm\,\Omega &=& \MyIm\,\left(\,dz^1\wedge dz^2 \right) . \nonumber
\end{eqnarray}

\noindent
The tangent planes of a generic two manifold in ${\mathbb{R}}^4$ are in one 
to one correspondence with the space of two planes in ${\mathbb{R}}^4$, 
$G(2,4)$, which can be identified with the manifold $S^2_{+}\times S^2_{-}$.
Each of the two spheres is embedded in the space of self- or anti-self-dual 
2-forms, respectively. If one identifies the sphere of complex structures 
on ${\mathbb{R}}^4$ with one of the $S^2_{\pm}$, say $S^2_{+}$ (the 
difference is only the choice of orientation of ${\mathbb{R}}^4$), 
the calibration condition 
restricts the remaining planes to $J_0\times S^2_{-}$. 
This reflects the fact, that the K\"ahler cycle is the one where the 
symplectic form becomes maximal. 
In principle one can compute the complex combination for each point in 
$S_{+}^2$. In section~\ref{SubSectionStringsOnM5} we are concerned with 
the following pencil of complex structures 
\begin{eqnarray}
    (a_1, a_2, a_3) ~=~ (\,-\,\sin\tilde\varphi,\,-\,\cos\tilde\varphi,\,0\,)
\end{eqnarray}  
with the complex coordinates 
\begin{eqnarray*}
   \tilde{z}^1 &=& q^1\,-\,i\,\left(\,
                                 \sin\tilde\varphi\cdot p^1\,+\,
                                 \cos\tilde\varphi\cdot q^2
                              \right)\\
   \tilde{z}^2 &=& p^2\,-\,i\,\left(\,
                                 \cos\tilde\varphi\cdot p^1\,-\,
                                 \sin\tilde\varphi\cdot q^2
                              \right).
\end{eqnarray*}
This could be seen as a rotation inside the $q^2-p^1$-plane.\\

\noindent
The projections of the K\"ahler form in the rotated complex coordinates 
$\tilde{z}^1,\,\tilde{z}^2$ to the old basis in eq.~(\ref{BasisSelfdual}) 
leads to the equations (\ref{Symplectic}) - (\ref{ImagTeil}).

\section{Definition of the commutator}
\label{AppendixGeomQuant}

\noindent
Here we collect all the details necessary to explain the transition 
from the Poisson bracket to the commutator of coordinates. 
The construction uses essentialy a symplectic structure $\omega$. 
In normal form it reads $\omega\,=\,\sum_i dq^i\wedge dp^i$. Here 
$q^i$ and $p^i$ are the conjugated variables. To define a  Poisson bracket 
we need the symplectic gradient $\sgrad g$ of a function $g$. It is 
simply defined by
\begin{eqnarray*}
       \sgrad g &=& \left(
                          \begin{array}{cc}
                             0 & \unity \\
                          -\unity & 0
                          \end{array}
                    \right)\cdot \grad g. 
\end{eqnarray*}
Now the Poisson bracket of two functions $g$ and $h$ is 
defined via $\omega$  by the formula: 
\begin{eqnarray*}
     \{g,h\} ~=~ -\,\omega(\, \sgrad g,\,\sgrad h\,). 
\end{eqnarray*}
For the symplectic structure above this leads to the standard relations
\begin{eqnarray}\label{Poissonklammer}
     \{x^i,p^{j}\} ~=~ \delta^{ij}. 
\end{eqnarray}
This is a realisation of the usual Heisenberg algebra 
${\mathcal{A}}_H$.\\
So far the construction is standard. Quantization can be performed by 
applying a construction called geometric quantization \cite{Woodhouse}. 
Given a symplectic manifold $(M,{\omega})$ it construct a linear map 
${\rho}:{\mathcal{A}} \longrightarrow {\mathcal{O}}$ from an algebra 
${\mathcal{A}}$ of functions on $M$ to the set of hermitian 
operators ${\mathcal{O}}$ on a Hilbert space\footnote
%%%%%%%%%%%%%%%%%%%%%%%%%%%%%%%%%%%%%%%%%%%%%%%%%%%%%%%%%%%%%%%%%%%%%%%%
% FOOTNOTE
%%%%%%%%%%%%%%%%%%%%%%%%%%%%%%%%%%%%%%%%%%%%%%%%%%%%%%%%%%%%%%%%%%%%%%%%
{
  The technical difficulty is the precise definition of the Hilbert 
  space. Usualy the space of functions on $M$ is to large. 
}
%%%%%%%%%%%%%%%%%%%%%%%%%%%%%%%%%%%%%%%%%%%%%%%%%%%%%%%%%%%%%%%%%%%%%%%%
${\mathcal{H}}(M)$, satisfying 
the following constraints: 
\begin{eqnarray}\label{GQ1} 
   \rho(1)\hspace{3ex} &=& \unity_{\mathcal{H}},{\rm if}\: 
                           1\in{\mathcal{A}}\\[0.5ex]
   \rho(F^\ast)\hspace{1ex}\; &=& \rho(F)^\dagger\label{GQ2}\\[0.5ex]
   [\rho(F),\rho(G)] &=& -i\hbar\rho(\{F,G\})  \;\; \forall \;\; F,G \in {\mathcal{A}}\label{GQ3}
\end{eqnarray}
On a patch $U_\alpha\subset M$ the map $\rho$ is given by 
\begin{eqnarray}\label{Prequant}
    \rho_\alpha(F) ~=~ -i\hbar\;\sgrad F \,+\, F \,-\, 
                       \theta_\alpha(\,\sgrad F\,)
\end{eqnarray}
with ${\theta}_\alpha\,=\,\sum_{k=1}^n p_kdq_k$ and 
$\omega\,=\,d\theta_\alpha$. The consistency of the construction leads 
to the Bohr-Sommerfeld quantization:
\begin{eqnarray}\label{BohrSommerfeld}
   \frac{1}{2\pi\hbar}\int\limits_{2-cycle}\omega ~\in~ {\mathbb{Z}}.
\end{eqnarray}
Appying the procedure outlined before to the case at hand by 
combining eq.~(\ref{Poissonklammer}) and eq.~(\ref{GQ3}) leads to 
\begin{eqnarray}\label{Kommutator}
    [\,\rho(x^i),\,\rho(p^j)\,] ~=~ -\,i\,\hbar\,\delta^{ij}.
\end{eqnarray}
with $\hbar$ a free constant (here $\hbar\,=\,1$).\\

\noindent
The formula eq.~(\ref{BohrSommerfeld}) is the first Chern number of the 
line bundle with curvature $\omega$. Its physical interpretation is the 
following. By the $S^1-$compactification the M2- and M5-branes are reduced 
to a fundamental string or D4-brane, respectively. Each of these branes 
carries a winding number, one of which classifies the magnetic the other 
one the electric charge of the system. 
The magnetic charge is precisely the number above. The electric charge 
is invisible in our geometric setup. The reason is quite simple. Since 
we are dealing only with a single electrically  charged object, there 
is no possibility to define what is meant by the minimal quantum of 
electrical charge.

\end{appendix}

%%%%%%%%%%%%%%%%%%%%%%%%%%%%%%%%%%%%%%%%%%%%%%%%%%%%%%%%%%%%%
%      Section: Acknowledgement  
%%%%%%%%%%%%%%%%%%%%%%%%%%%%%%%%%%%%%%%%%%%%%%%%%%%%%%%%%%%%%

\vskip0.5cm
\noindent{\bf Acknowledgements:}

\noindent 
We want to thank Ralph Blumenhagen for useful discussions.
The work was partly supported by the EC contract
HPRN-CT-2000-00131. B.K. also wants to thank
the Studienstiftung des deutschen Volkes. 

%%%%%%%%%%%%%%%%%%%%%%%%%%%%%%%%%%%%%%%%%%%%%%%%%%%%%%%%%%%%%
%   BIBLIOGRAPHY
%%%%%%%%%%%%%%%%%%%%%%%%%%%%%%%%%%%%%%%%%%%%%%%%%%%%%%%%%%%%%

\bibliographystyle{utphys}
%\bibliography{starreport12.bbl}

\begingroup\raggedright\endgroup

\end{document}